\documentclass{PoS}
\usepackage{graphicx}

\newcommand{\eqref}[1]{(\ref{#1})}

\newcommand{\nn}{\nonumber\\}
\newcommand{\ph}{\varphi}

\newcommand{\ep}{\varepsilon}

\newcommand{\pint}[2]{{\int\!\frac{d^{#1}#2}{(2\pi)^{#1}}\,}}

\newcommand{\MSbar}{{\ensuremath{\overline{\mathrm{MS}}}}}

\renewcommand{\c}[1]{{\cal{#1}}}

\newcommand{\tehat}{\ensuremath{\quad\Rightarrow\quad}}

\title{Reconciling resummation and renormalization }
\ShortTitle{Reconciling resummation and renormalization }
\author{\speaker{Antal Jakovac}\\
  BME, Institute of Physics, Budafoki \'ut 8, H-1111 Budapest , Hungary\\
  E-mail: \email{Antal.Jakovac@cern.ch}}

\abstract{In the naive form of most resummations we get into
  conflict with order-by-order renormalization. We present a
  method that is capable to ensure UV consistency of any
  resummations satisfying certain conditions. The method is
  based on the observation that resummation is equivalent with a
  calculation in an adequate perturbation scheme, followed by a
  renormalization scheme changing. This framework works both in
  static and momentum-dependent cases. In particular it is
  possible to establish finite infrared 2PI resummation. }

 \FullConference{29th Johns Hopkins Workshop on current problems in
   particle theory: strong matter in the heavens\\ 
   1-3 August\\
   Budapest}

\begin{document}

\section{Introduction}

In a theory with small coupling constants we expect that perturbation
theory works well; in fact what we understand under this sentence is
that renormalized perturbation theory with some standard
renormalizations scheme (eg. \MSbar\ or on-mass-shell scheme) provides
small corrections as compared to the leading order. When this approach
fails, we tend to speak about the failure of perturbation theory
itself. This failure can sometimes be associated to certain class of
diagrams in the chosen scheme. Then we have the hope that after
resummation of this subset of diagrams we can still give analytic
predictions for the observable in question. We can give several
examples when we need resummation: self-energy resummations
(Schwinger-Dyson equation) is needed to determine mass shift, daisy
resummation \cite{DolanJackiw} is needed at high temperatures in a
scalar field theory, HTL resummation \cite{BrPis} should be used in
gauge theories at high temperatures. In 2PI and higher point
irreducible resummation we can work with exact propagators/vertices
\cite{Berges_sum}. And also the renormalization group flow defines a
set of resummed theories parameterized by the renormalization scale
\cite{Collins}.

In most cases resummations are designed to solve some infrared (IR)
problems, ie. it resums diagrams that are the most sensitive in the IR
regime. The corresponding counterterm diagrams that are needed to
ensure finiteness of the perturbation theory at high momenta, are often
neglected. This has a consequence that ultraviolet (UV) consistency
becomes a serious problem in the resummed theories. The problem
becomes even deeper in case of momentum dependent resummations, like
the 2PI resummation. There are different approaches published recently
to overcome this difficulty, primarily in the static and 2PI
resummation case \cite{recent}.

In the present paper we try to describe an approach different from the
above, mainly diagrammatic methods. The idea is that the only reliable
analytic method to treat UV divergences is the renormalized
perturbation theory \cite{Collins}. In some way all the consistent
resummations have to be linked to a specific scheme. The task is to
find this link for each specific resummation method. 

To understand the relation between the resummation methods and
renormalization schemes we recall that renormalized perturbation
theory has a large freedom in choosing the finite parts of the
counterterms. We can use this freedom to choose finite parts such a
way that mostly reduces the IR sensitivity of the system. The
so-defined scheme will depend on the environmental parameters (like
temperature). In order to have results in a reference scheme (eg.
\MSbar) we must perform a matching between the parameters of the
perturbation scheme and the reference scheme, using the requirement
that the \emph{bare Lagrangian} are the same. Both being
renormalization schemes, this is sufficient to match all of the
observables to the given order in perturbation theory \cite{Collins}.
The difference of the \emph{higher order terms} is the resummation.
The idea of fitting the perturbation theory to the environment was
used already in \cite{envfriend} where the authors tuned the
renormalization scale appropriately.

This strategy will be described in this contribution in more detail.
First we examine the problem in the simplest static mass resummation
case, and we show how the method sketched above will solve the UV
consistency problem, in principle and in the $\Phi^4$ theory. For a
more elaborated description of these section cf. \cite{JakovacSzep}.
Then we change to the momentum dependent case, where basically the
same method works, but we have to take care some details. As a special
application we present how a 2PI resummation can be represented by a
scheme, and how can it solve the renormalizability problem. Finally we
give a short summary of the method.

\section{Thermal mass resummation -- the naive method}

A nice and well-known example of a system that needs resummation is
the high temperature $\Phi^4$ theory. In this model, as was shown by
Dolan and Jackiw \cite{DolanJackiw}, at $n$-th order of perturbation
theory we obtain a mass correction of order $\lambda T^2 (\lambda
T^2/m^2)^{n-1}$, which is a growing function of the coupling constant
if $T\gg m$. In this case the class of daisy diagrams was the adequate
subset which was needed to sum up. To facilitate the treatment, the
same effect could have been achieved with the thermal counterterm
method \cite{thermalcount}, where we subtracted from and added to the
Lagrangian the same term, but treating them at different loop order.
In this way it was possible to change the tree level mass without the
change of the Lagrangian. If we denote the free mass-squared by $m^2$
and the resummed mass by $M_T^2=m^2+\Delta M_T^2$, then the mass terms
of the Lagrangian will be written as
\begin{equation}
  -\c L_\mathrm{mass} =\underbrace{\frac{M_T^2}2\ph^2
      \rule[-1.3em]{0em}{3em}}_{\mbox{\small tree level}} +
    \underbrace{\left(- \frac{M_T^2-m^2}2\ph^2 +\frac{\delta
          m^2}2\ph^2\right)}_{\mbox{\small one loop}}.
\end{equation}
To determine the ideal value of the added-subtracted mass one should
use an additional requirement, for example that the one-loop level
mass term is just $M_T^2$. Denoting the complete unresummed one-loop
self energy at momentum $k$ and with $M^2$ mass-squared on the
internal lines by $\Pi(k;M^2)$ we find
\begin{equation}
  \label{gapeq}
  M_T^2 = m^2 + \Pi(k=0,M_T^2).
\end{equation}
This is an implicit (gap) equation for the mass $M_T^2$ as a function
of $m^2$. In other models, or for other infrared (IR) problems other
type of resummations proved to be useful (like super-daisy, HTL, 2PI,
RG etc.)

This appealing method has, however, a severe drawback: taking it
really seriously it provides ultraviolet (UV) divergent result. The
symbol $\Pi$ in \eqref{gapeq}, in fact has contributions from the
tadpole diagram and the mass counterterm. The tadpole contribution,
according to \eqref{gapeq}, has to be computed with internal
propagators with $M_T^2$ mass-squared; it reads
\begin{equation}
  \label{TB}
  T_B(M_T^2) = \frac{M_T^2}{16\pi^2} \left[-\frac1\ep +\gamma_E-1
    +\ln\frac{M_T^2}{4\pi\mu^2} \right] +
  \frac1{2\pi^2}\int\limits_M^\infty \!d\omega\,\sqrt{\omega^2-M_T^2}
  \,n(\omega).
\end{equation}
The mass counterterm, on the other hand, has a value fixed by the
renormalization scheme; if we use $\MSbar$ then we have
\begin{equation}
  \label{deltam2MSbar}
   \delta m^2 = - \frac{\lambda m^2}{32\pi^2}\, \left[-\frac1\ep
     +\gamma_E-1 +\ln\frac{M_T^2}{4\pi\mu^2} \right]
\end{equation}
The self-energy therefore
\begin{equation}
  \Pi(k=0,M_T^2) = \frac \lambda 2\, T_B(M_T^2) + \delta m^2 =
  -\frac{\lambda(M_T^2- m^2)}{32\pi^2}\;\frac1\ep  +\dots
\end{equation}
is UV divergent.

\section{Treating UV divergences in perturbation theory}

The main reason for this divergence is that the finiteness of
perturbation theory depends on a very sensitive balance between the
counterterms and 2PI diagrams \cite{Collins}. In fact only a small
subclass of all conceivable perturbation theories can fulfill all the
requirements to provide order-by-order finiteness: these are the
renormalized perturbation series of a renormalizable theory. So when
UV finiteness is a crucial issue, then we must remain within this
subclass, we must use renormalized perturbation theory for
resummation, too.

Fortunately this subclass is rather wide, as we can freely choose the
finite part of all the counterterms at all orders. All choices yield
different renormalized perturbation series; ie. they provide different
results at any fixed order as a function of the renormalized
parameters of the Lagrangian. One is common in all of these results:
they are all finite. These span the space of ``perturbatively
reachable'' domain in a given theory. We should speak about
``non-perturbative effect'' only if some phenomenon lies outside of
this domain.

As the example of the thermal mass resummation indicated, a generic
perturbation theory has a very small convergence radius (in the weak
``asymptotic convergence'' sense) because of IR sensitivity. Only
schemes well adapted to the environment \cite{envfriend} show good
convergence properties. To give a well-known example: in the $\Phi^4$
theory with negative mass-squared we do not work in the original free
Hilbert space, since there the masses are imaginary, instead we adapt
the perturbation theory to the (expected) vacuum properties and use a
Hilbert space built on the spontaneously broken vacuum.

Usually we want to compare result coming from different environments:
for example we want to know the thermal properties as a function of
the zero temperature observables. Different environments need
different schemes, whose results are not directly comparable. This is
because different choices of the finite parts formally yield different
bare Lagrangian and so different physics. But the bare Lagrangian is
completely determined by the bare parameters (as bare masses,
couplings, wave function renormalizations), so if we require that the
bare parameters are equal we can hold the physics constant in the two
schemes. Expressing as function of renormalized parameters in each
specific schemes, the equality of bare parameters impose relations
between the renormalized parameters of different schemes. For example
in $\Phi^4$ theory, where we have 3 renormalized parameters ($Z$ wave
function renormalization constant, $m^2$ mass-squared and $\lambda$
coupling constant), we will have the following relations
\begin{equation}
  Z_A = Z_B,\qquad  m_A^2+\delta m_A^2 =  m_B^2+\delta m_B^2, \qquad
  \lambda_A+\delta\lambda_A = \lambda_B+\delta\lambda_B.
\end{equation}
If two schemes are related in this way, then there different results
for a given observable can be considered as a \emph{resummation
  effect}, since at infinite loop order they both yield the same,
exact result.

\section{Example: thermal mass resummation in $\Phi^4$ theory}

Using the ideas above we can work out the thermal mass resummation in
a consistent way. In \emph{any} perturbation schemes the one-loop
self-energy reads
\begin{equation}
  \Pi(k=0) = \frac\lambda 2 T_B(m^2) + \delta m_1^2,
\end{equation}
where $T_B$ was defined in \eqref{TB}. In \MSbar\ scheme we choose
\eqref{deltam2MSbar} for the mass counterterm. However, only the
divergent part is really fixed by the condition of renormalizability,
different schemes can use different finite parts. A specific choice
can be:
\begin{equation}
  \delta m_{1,\mathrm{res}}^2 = - \frac\lambda 2 T_B(m^2).
\end{equation}
This scheme depend on the temperature, but only through the finite
parts. At one hand this is allowed mathematically, on the other hand
this is expected: the environment, we have to be adapted to, is
represented by the temperature. In this \emph{res} scheme the one loop
self energy is zero
\begin{equation}
  \Pi_{\mathrm{res}}(k=0) = 0,
\end{equation}
so it is in fact a finite temperature mass-shell scheme: the complete
self energy is just the mass.

If we use \MSbar\ at zero temperature and \emph{res} scheme at finite
temperature we have to ensure that they describe the same physics, ie.
they stem from the same bare Lagrangian. In this simple example this
requirement reduces a relation between the renormalized mass values in
the two schemes. Expressing the bare mass to order $\lambda$ in both
cases, we find
\begin{equation}
  Z^2 m_\mathrm{bare}^2 = m_\mathrm{res}^2 + \delta m_{1,\mathrm{res}}^2 =
  m_\MSbar^2 + \delta m_\MSbar^2 \tehat m_\MSbar^2 = m_\mathrm{res}^2
  - T_{B,\MSbar}(m_\mathrm{res}^2),
\end{equation}
where the last symbol means
\begin{equation}
  T_{B,\MSbar}(m_\mathrm{res}^2) = \frac{\lambda
    m_\mathrm{res}^2}{32\pi^2} \ln\frac{m_\mathrm{res}^2}{\mu^2} +
  \frac\lambda{4\pi^2}\int\limits_{m_\mathrm{res}}^\infty
  \!d\omega\,\sqrt{\omega^2-m_\mathrm{res}^2} \,n(\omega),
\end{equation}
which is nothing but the tadpole diagram renormalized in \MSbar\
scheme, evaluated at $m^2=m_\mathrm{res}^2$ point. So, in fact, the
gap equation \eqref{gapeq} is true in the sense that $\Pi$ is the
renormalized self energy correction.

\section{Momentum dependence}

What was said so far applies for the momentum-independent
resummation. In many cases, however, this is not enough to fully
diminish IR sensitivity from the system. In these cases we have to
apply momentum dependent resummations.

The way we have adapted the renormalization scheme to the environment
was the proper choice of finite parts of the counterterms. Momentum
dependent resummations therefore imply usage of momentum dependent
finite parts. Formally this is feasible, the only question is that
whether in this way we do not spoil renormalizability.

Let us concentrate on the mass resummation in this proceedings, the
more elaborated complete discussion will be published elsewhere
\cite{inprep}. We now choose $\delta m^2(k)$, momentum dependent
counterterm. In this case the matching to a reference scheme, eg.
\MSbar, yields the condition
\begin{equation}
  Z^2m_\mathrm{bare}^2 = m_\MSbar^2 +   \delta m_\MSbar^2 = m^2 +
  \delta m^2(k).
\end{equation}
This enforces to work with momentum dependent tree level mass from the
beginning, ie. $m^2=m^2(k)$.

We can maintain renormalizability, provided we satisfy some
requirements. First the divergent part of the counterterm should be
unique, ie. the same as, say, for the \MSbar\ case. Secondly, the
momentum dependence of the tree level mass should not generate new
divergences. We will assume that for asymptotically large momentum the
tree level mass behaves as
\begin{equation}
  m^2(k) = m^2_R + \c O(k^{-\gamma}).
\end{equation}
The most singular diagram is the tadpole. If there are no new
divergences in the tadpole, then there are no new divergences in any
other diagrams. With momentum dependent mass the value of the tadpole
reads
\begin{eqnarray}
  \pint4p \frac1{p^2-m^2(p)} && = \pint4p \frac1{p^2-m_R^2
  -\c O\left(p^{-\gamma}\right)}=\nn &&=  \pint4p\left[
  \frac1{p^2-m_R^2} + \c O\left(\frac{p^{-\gamma}} {(p^2-m_R^2)^2}
  \right)\right].
\end{eqnarray}
The first term is the usual tadpole contribution. The second term is
finite if $\gamma>0$. So if the momentum dependent mass approaches its
limiting value as a power law, then the divergence structure remain
untouched.

\section{The 2PI resummation}

Let us write the self energy as 
\begin{equation}
  \Pi(k) = \delta m^2 + \bar\Pi(k,m^2).
\end{equation}
In the mass-shell scheme we choose $\delta m^2 = -\bar \Pi(k=m^2)$ at
zero temperature. We have seen that if we apply the same prescription
at finite temperature, it leads to thermal mass resummation. Going on
with this idea we may try to choose
\begin{equation}
  \delta m^2(k) = -\bar\Pi(k,m^2(k))
\end{equation}
at \emph{any momenta}. This has the consequence that $\Pi(k)=0$, ie.
there is no self-energy correction whatsoever in this scheme! As a
resummation, therefore, it provides self-energy correction resummation
(also in the internal lines!), which is just the 2PI resummation. So
we will call this scheme as 2PI scheme.

If we compare with the \MSbar\ scheme the requirement of the constant
physics reads
\begin{equation}
  m^2(k) - \bar\Pi(k,m^2(k)) = m_\MSbar^2 + \delta m^2_\MSbar .
\end{equation}
Since the divergence structure of $-\bar\Pi(k,m^2(k))$ is the same
as that of $\delta m_\MSbar^2$ the expression $\bar\Pi(k,m^2(k))+\delta
m^2_\MSbar$ is finite, and diagrammatically it contains no self-energy
correction in the internal lines; therefore we will call it $\bar
\Pi_{2PI}$ since its meaning is really the self-energy renormalized in
the 2PI scheme.

In the language of the propagator $G^{-1}(k) = k^2 -m^2$ the above
relation can be written as
\begin{equation}
  G_{2PI}^{-1}(k) = G_\MSbar^{-1}(k) + \bar\Pi_{2PI}(k, G_{2PI}(k)).
\end{equation}
If we identify $G_\MSbar$ as the ``free propagator'' and $G_{2PI}$ as
the ``resummed propagator'', then this equation is exactly the basic
relation of the 2PI resummation \cite{Berges_sum}.

But in the light of renormalizability the above procedure is not
satisfactory. In the perturbative evaluation of $\bar \Pi(k)$ we will
encounter terms like $k^2\ln k$ or $m^2\ln k$, which are not allowed
contributions to $\delta m^2(k)$. Therefore the naive 2PI approach is
not renormalizable -- as it is well known in the 2PI literature
\cite{Berges_sum}.

It is possible to give a solution to the problem of 2PI
renormalizability in the present context. We are not enforced now to
choose exactly the $\bar\Pi$ self energy; after all, resummation is
needed by IR sensitivity, ie. a low momenta phenomenon. It is enough
to perform 2PI resummation in this domain, too. We can, therefore,
simply omit the terms that high momentum terms that would violate
renormalizability. We can do it with some (smooth) cutoff function
$\Theta(x)$, and choose, for example
\begin{equation}
  \delta m^2(k) = - \Theta(k/\Lambda) \bar\Pi(k,m^2(k)) -
  (1-\Theta(k/\Lambda)) \delta m_\MSbar^2.
\end{equation}
For another solution we realize that since the problem comes from
asymptotically large momenta this is insensitive to the environment;
in particular they appear in the same way at zero temperature, and in
\MSbar\ scheme. If we subtract the zero temperature perturbative self
energy renormalized in \MSbar, the rest is already appropriate to play
the role of a momentum dependent mass counterterm. We should be
careful, however, since asymptotically divergent contributions may
come from sub-diagrams, and so the subtraction has to be repeated to
every sub-diagram, in the spirit of the forest formula. After these
subtractions we can write
\begin{equation}
  \delta m^2(k) = - \bar\Pi_{asympt.\ subtr.}(k,m^2(k)).
\end{equation}

\section{Summary}

To summarize the content of this contribution we repeat that we must
not resum IR sensitive diagrams without respect of the UV consistency,
otherwise we run into divergences even in the most simple cases. The
perturbative method that is capable to ensure UV finiteness of a
renormalizable theory, is the renormalized perturbation theory. We
have a freedom in the choice of the finite parts of the counterterms
(which defines the scheme), and so we have the possibility to adapt
the finite parts to the environment. To compare result coming from
different schemes we must ensure that the physics is the same;
formally this means the requirement to keep the bare Lagrangian
constant.

This line of thought can be continued and apply to the momentum
dependent resummations, since formally the finite parts of the
counterterms can be momentum dependent. To be consistent we must allow
the appearance of momentum dependent renormalized parameters, too.
Renormalizability can be maintained, if the momentum dependence is
soft enough in the asymptotic momentum domain: in case of the mass
term, for example, $m^2(k) = m_R^2 + \c O(k^{-\gamma})$ momentum
dependence is allowed for large momenta, where $\gamma>0$.

In this way we can reproduce the 2PI resummation with momentum
dependent schemes. The naive definition, however, turn out to be
non-renormalizable -- just like the usual 2PI resummation. In this
context the problem of renormalizability can be solved by omitting the
problematic terms in the asymptotic momentum region. In this way we
should abandon the complete self-energy (2PI) resummation, but we can
still maintain the 2PI resummation in the infrared regime.

\section*{Acknowledgment}

The author thanks for A. Patk\'os, U. Reinosa and Zs. Sz\'ep for
discussions. Support by the Hungarian National Research Fund OTKA
(F043465), by the Hungarian Research and Technological Innovation
Fund, and by the Croatian Ministry of Science, Education and Sports is
gratefully acknowledged.

\end{document}